\documentstyle[12pt,psfig]{article}
\catcode`\@=11
\def\marginnote#1{}
\newcount\hour
\newcount\minute
\newtoks\amorpm
\hour=\time\divide\hour by60
\minute=\time{\multiply\hour by60 \global\advance\minute by-
\hour}
\edef\standardtime{{\ifnum\hour<12 \global\amorpm={am}%
    \else\global\amorpm={pm}\advance\hour by-12 \fi
    \ifnum\hour=0 \hour=12 \fi
    \number\hour:\ifnum\minute<100\fi\number\minute\the\amorpm}}
\edef\militarytime{\number\hour:\ifnum\minute<100\fi\number\minute}
\def\draftlabel#1{{\@bsphack\if@filesw {\let\thepage\relax
  \xdef\@gtempa{\write\@auxout{\string
    \newlabel{#1}{{\@currentlabel}{\thepage}}}}}\@gtempa
    \if@nobreak \ifvmode\nobreak\fi\fi\fi\@esphack}
     \gdef\@eqnlabel{#1}}
\def\@eqnlabel{}
\def\@vacuum{}
\def\draftmarginnote#1{\marginpar{\raggedright\scriptsize\tt#1}}
\def\draft{\oddsidemargin -.5truein
        \def\@oddfoot{\sl preliminary draft \hfil
        \rm\thepage\hfil\sl\today\quad\militarytime}
        \let\@evenfoot\@oddfoot \overfullrule 3pt
        \let\label=\draftlabel
        \let\marginnote=\draftmarginnote
   
\def\@eqnnum{(\theequation)\rlap{\kern\marginparsep\tt\@eqnlabel}%
\global\let\@eqnlabel\@vacuum}  }
\def\preprint{\twocolumn\sloppy\flushbottom\parindent 1em
        \leftmargini 2em\leftmarginv .5em\leftmarginvi .5em
        \oddsidemargin -.5in    \evensidemargin -.5in
        \columnsep 15mm \footheight 0pt
        \textwidth 250min      \topmargin  -.4in
        \headheight 12pt \topskip .4in
        \textheight 175mm
        \footskip 0pt
        
\def\@oddhead{\thepage\hfil\addtocounter{page}{1}\thepage}
        \let\@evenhead\@oddhead \def\@oddfoot{} \def\@evenfoot{} 
}
\def\titlepage{\@restonecolfalse\if@twocolumn\@restonecoltrue\onecolumn
     \else \newpage \fi \thispagestyle{empty}\c@page\z@
        \def\thefootnote{\fnsymbol{footnote}} }
\def\endtitlepage{\if@restonecol\twocolumn \else  \fi
        \def\thefootnote{\arabic{footnote}}
        \setcounter{footnote}{0}}  %\c@footnote\z@ }
\catcode`@=12
\relax
\def\be{\begin{equation}}
\def\ee{\end{equation}}
\def\bea{\begin{eqnarray}}
\def\eea{\end{eqnarray}}
\def\simlt{\stackrel{<}{{}_\sim}}
\def\simgt{\stackrel{>}{{}_\sim}}

\def\NPB#1#2#3{{\it Nucl.~Phys.} {\bf{B#1}} (19#2) #3}
\def\PLB#1#2#3{{\it Phys.~Lett.} {\bf{B#1}} (19#2) #3}
\def\PRD#1#2#3{{\it Phys.~Rev.} {\bf{D#1}} (19#2) #3}

\def\ZPC#1#2#3{{\it Z.~Phys.} {\bf C#1} (19#2) #3}
\def\PTP#1#2#3{{\it Prog.~Theor.~Phys.} {\bf#1}  (19#2) #3}

\def\PR#1#2#3{{\it Phys.~Rep.} {\bf#1} (19#2) #3}

\def\msbar{\overline{\rm MS}}

\relax
\begin{document}
\topmargin-2.5cm
%\draft
%\preprint
%
\begin{titlepage}
\begin{flushright}
SCIPP-96-04\\
DESY 96-021 \\
IEM-FT-123/96 \\
hep--ph/9603227 \\
\end{flushright}
\vskip 0.3in
\begin{center}{\large\bf 
STANDARD MODEL STABILITY BOUNDS FOR  \\ 
\vspace{.25cm} NEW PHYSICS WITHIN LHC REACH
\footnote{Work supported in part by 
the European Union (contract CHRX-CT92-0004) and
CICYT of Spain 
(contract AEN95-0195).}  }
\vskip .5cm
{\bf J.A. Casas \footnote{On leave of absence from 
Instituto de Estructura de la Materia (CSIC), Serrano 123
28006-Madrid (Spain).} }\\
Santa Cruz Institute for Particle Physics, University of
California,\\ Santa Cruz, CA 95064, USA\\
\vskip.25cm
{\bf J.R. Espinosa \footnote{Supported by
Alexander-von-Humboldt Stiftung.}
}\\
Deutsches Elektronen-Synchrotron DESY, Hamburg, Germany \\
and \\ 
{\bf M. Quir\'os}\\
Instituto de Estructura de la Materia (CSIC), 
Serrano 123 28006-Madrid, Spain\\
\end{center}
\vskip.5cm
\begin{center}
{\bf Abstract}
\end{center}
\begin{quote}
We analyse the stability lower bounds on the Standard Model Higgs 
mass by carefully controlling the scale independence of the effective 
potential. We include resummed leading and next-to-leading-log corrections, 
and physical pole masses for the Higgs boson, $M_H$, and the 
top-quark, $M_t$. Particular attention is devoted to the cases where
the scale of new physics $\Lambda$ is within LHC reach, i.e. 
$\Lambda\leq 10$ TeV, which have been the object of recent
controversial results. We clarify the origin of discrepancies
and confirm our earlier results within the error of our previous
estimate. In particular for $\Lambda=1$ TeV we find that 
$$M_H[GeV]> 52+0.64\ (M_t[GeV]-175)
-0.50\ \frac{\alpha_s(M_Z)-0.118}{0.006}.$$ 
For fixed values of $M_t$ and $\alpha_s(M_Z)$, the error from 
higher effects, as the lack of exact scale invariance of the
effective potential and higher-order radiative corrections, 
is conservatively estimated to be $\simlt 5$ GeV.

\end{quote}
\vskip1.cm

\begin{flushleft}
IEM-FT-123/96\\
February 1996 \\
\end{flushleft}

\end{titlepage}
\setcounter{footnote}{0}
\setcounter{page}{0}
\newpage
%
% BODY
\newpage
%{\bf 1.}

\section{Introduction}
The requirement of vacuum stability in the Standard Model (SM)
imposes a severe lower bound on the Higgs mass, $M_H$, which depends on
the precise value of the top-quark mass, $M_t$, and on the scale
$\Lambda$ beyond which the SM is no longer valid \cite{bounds}.
The relationship between the scale of new physics $M$ [the mass of new
particles or resonances] which can stabilize the effective
potential, and the instability scale $\Lambda$ has recently been
studied in Ref.~\cite{HS}. The conclusion is that $M$ can be a
few times $\Lambda$ provided that new physics is strongly
coupled. However, if new physics is weakly coupled, consistently
with the idea of Grand [or String] Unification, then we 
expect that~\footnote{A good example is the case of the
Minimal Supersymmetric Standard Model where the scalar particles
that can stabilize the effective potential are the third
generation squarks, with a multiplicity N=12 and a coupling [in
the notation of Ref.~\cite{HS}] $\delta=h_t^2/2$. For
$m_t\simlt 190$ GeV we obtain $\delta\simlt 0.6$, which
corresponds from Fig.~2 of \cite{HS} to $M\simlt\Lambda$.}
$\Lambda\sim M$. In this case it is of the utmost interest to study 
scenarios with $\Lambda$ scales below roughly $10$ TeV since they 
correspond to cases where new (weakly interacting) physics should be 
produced at LHC. This is stressed by the 
fact that the typical lower bounds on the Higgs mass in these scenarios 
lie precisely around the range accessible to LEP2.

Updated stability bounds have been presented in
Refs.~\cite{AI,CEQ}, AI and CEQ bounds, respectively. Both papers
agree for large values of $\Lambda$
[i.e. $\Lambda=M_P$ or even several orders of magnitude smaller, depending
on the cases], while they differ substantially for low values of $\Lambda$. 
In particular, for $\alpha_s(M_Z)=0.118$,
$M_t=175$ GeV and 
$\Lambda=10^{19}$~GeV, AI quote
$M_H>137$ GeV and CEQ $M_H>133$ GeV, a difference well within
the theoretical errors. However, for the same value of $M_t$ and
$\Lambda=1$ TeV, AI quote $M_H>73$ GeV while CEQ give a much
lower bound $M_H>55$ GeV, a substantial difference which cannot
be absorbed in the quoted errors of the two calculations. Notice that
the region of discrepancy lies precisely in the 
region of interest for LHC prospects as mentioned above.

Very recently, the lower stability bounds on the SM
Higgs mass have been reconsidered~\cite{W1,W2}. The discrepancy 
between AI and CEQ results has been claimed~\cite{W2} to be explained by 
the non inclusion in Ref.~\cite{CEQ} of the Higgs-Yukawa sector 
contributions (more precisely, tadpole contributions) in the 
relation between the top-quark pole and $\msbar$ running masses. 

In this letter we will show that the claim in Ref.~\cite{W2} is
incorrect because finite electroweak tadpoles are automatically
included in the treatment of Ref.~\cite{CEQ}. As we will see, the
discrepancy between AI and CEQ bounds for low values of
$\Lambda$ can be traced back to the more accurate description of the
effective potential in Ref.~\cite{CEQ}, in particular with the inclusion 
of one-loop effects. For large values of $\Lambda$, which correspond 
to the region of maximum concern in Refs.~\cite{AI,CEQ}, these
effects are in fact negligible and both calculations are in agreement.
However, these one-loop effects become relatively
important for low values of the instability scale and, by
removing them, we will precisely recover the AI bounds. Moreover, a
detailed analysis of the scale independence of the effective
potential provides a bound somewhat lower than in our previous
analysis. This makes, for the same value of $M_t$ and $\Lambda$
as above, the lower bound to decrease to $\sim 52$ GeV. This 
procedure will also allow
a reliable estimate of the various theoretical errors (coming from 
the lack of scale invariance of the effective potential, the
non-considered two-loop corrections, the gauge and renormalization scheme
dependence of the result, etc.) involved in the calculation.
All these precise estimates will be essential if a
Higgs boson with Standard Model properties is found at LEP2 (with a mass 
$\simlt$ 90 GeV) and will concern the need of new physics 
at LHC. 

\section{Higgs and top masses in the standard vacuum}
It has been shown~\cite{effpot} that the one-loop effective
potential improved by two-loop renormalization group equations
resums the next-to-leading-log contributions. To this order of
approximation, the SM effective potential can be written in the
't Hooft-Landau gauge and the $\msbar$ renormalization scheme as
\be
\label{potential}
V_{\rm eff}=V_0+V_1, 
\ee
where the tree-level, $V_0$, and one-loop,
$V_1$, terms are given by:
\be
\label{treelevel}
V_0=-\frac{1}{2}m^2(t)\phi^2(t)+\frac{1}{8}\lambda(t)\phi^4(t),
\ee
\be
\label{oneloop}
V_1=\sum_i\frac{n_i}{64\pi^2}M_i^4(\phi)\left[
\log\frac{M_i^2(\phi)}{\mu^2(t)}-C_i\right]
\ +\ \Omega(t),\ee
with~\footnote{The contribution from Higgs and
Goldstone bosons can be easily incorporated, though it is
numerically irrelevant as we have checked.} $i=W,Z,t$,
$M_i^2=\kappa_i\phi^2(t)$, and
$$ C_W=C_Z=\frac{5}{6},\ \ C_t=\frac{3}{2}, $$

$$ n_W=6,\ \ n_Z=3,\ \ n_t=-12, $$

$$ \kappa_W=\frac{1}{4}g^2(t),\ \
\kappa_Z=\frac{1}{4}[g^2(t)+g'^2(t)],\ \
\kappa_t=\frac{1}{2}h^2(t). $$

\noindent
In the previous expressions the parameters $\lambda(t)$ and
$m(t)$ are the SM quartic coupling and mass, and $g(t)$, $g'(t)$
and $h(t)$ are the $SU(2)$, $U(1)$ and top Yukawa couplings,
respectively. The running of the Higgs field is 
$\phi(t)=\xi(t)\phi_c$, $\phi_c$ being the classical field, and 
$\xi(t)=\exp\{-\int_0^t\gamma(t')dt'\}$ where $\gamma(t)$ is the
Higgs anomalous dimension. The scale $\mu(t)$ is related
to the running parameter $t$ by $\mu(t)=M_Z\exp(t)$. Finally,
$\Omega(t)$ is the (field-independent) one-loop contribution to
the cosmological constant [in particular we set it to $\Omega(t=0)=0$], 
which as we will see is irrelevant for the results of the present work.

{}From here on the procedure to fix the standard
electroweak minimum and the pole masses for the top-quark and
the Higgs boson is that specified in Refs.~\cite{CEQ,CEQR}.
For the sake of the discussion we will summarize the main points
here. The scale independence of the effective potential allows
fixing the renormalization scale $\mu(t)$ at will for different
values of the field~\cite{effpot}. Actually, the scale-invariance
properties of $V$ permit to perform the substitution either before
or after taking the derivative $\partial^n V/\partial \phi^n$, with
equivalent results \cite{CEQR}, which in turn allows to ignore
$\Omega$ for this task. On the other hand, although the whole effective
potential is scale-invariant, the one-loop approximation is not. Therefore,
one needs a criterion to choose the appropriate renormalization scale
in the previous equations. As was shown in \cite{CEQR} a sensible 
criterion is to choose as the optimal scale the value $\mu^*=\mu(t^*)$
where the potential is more scale-independent. This issue was carefully
examined in Ref.~\cite{CEQR}, where $\mu(t^*)$ was shown to be close to the 
top mass [its detailed value, however, is not very important because of
the high degree of scale independence of the one-loop potential
around $\mu(t^*)$]. Then, we minimize the potential
(\ref{potential}) at the scale $\mu(t^*)$. For the sake of the
discussion of the tadpole contribution, we next consider two (equivalent)
ways to do this. The first one is to define the tree-level 
vacuum expectation value (VEV) by means of the condition
$\left.\partial V_0(\phi,t^*)/\partial \phi\right|_{\phi=
\langle\phi(t^*)\rangle_0}=0$, i.e.
\be
\label{treevev}
\langle\phi(t^*)\rangle_0^2=\frac{2m^2(t^*)}{\lambda(t^*)}.
\ee
In this case one-loop corrections from 
(\ref{oneloop}) shift the VEV (\ref{treevev}) as
\be
\label{shift}
\langle\phi(t^*)\rangle^2=\langle\phi(t^*)\rangle_0^2+\delta\phi^2,
\ee
where (leaving apart gauge corrections)
\be
\label{deltaphi}
\delta\phi^2=-\frac{2}{\lambda\langle\phi\rangle}\frac{\partial 
V_1}{\partial\phi}\sim \frac{h_t^4}{\lambda}\langle\phi\rangle^2 ,
\ee 
which amounts diagrammatically to the contribution of the
one-loop tadpoles. This correction is large for low Higgs masses and has 
to be taken into account when relating $\langle\phi(t^*)\rangle^2_0$ to 
physical observables like $G_{\mu}$.

Another possibility~\cite{CEQ} is to define the one-loop VEV by means of
the condition $\left.\partial V_{\rm eff}(\phi,t^*)/\partial 
\phi\right|_{\phi=\langle\phi(t^*)\rangle}=0$, which can be expressed as
\be
\label{onevev}
m^2(t^*)=\frac{1}{2}\lambda(t^*)\langle\phi(t^*)\rangle^2
-\frac{3}{16\pi^2}h_t^4(t^*)\langle\phi(t^*)\rangle^2
\left[ \log\frac{h_t^2(t^*)\langle\phi(t^*)\rangle^2}
{2\mu^2(t^*)}-1 \right]+\cdots
\ee
where the ellipsis refers to gauge corrections.
Diagrammatically this procedure amounts \cite{HK,BW} to a
cancellation between the bare one-point vertex and the one-loop
tadpole contribution. In other words, tadpoles are absorbed in
the one-loop VEV and will never appear (except if we desired
to compute quantities from $V_0$). This is the procedure
followed in Ref.~\cite{CEQ} and the procedure we will adopt
here. We now impose the physical condition on the VEV
at $t=t^*$:
\be
\label{VEV}
\langle\phi(t^*)\rangle=\xi(t^*)\ v,
\ee
where $v=(\sqrt{2}G_{\mu})^{-1/2}=246.22$ GeV. 

Then the running of $m^2_H\equiv\partial^2V_{\rm
eff}/\partial\phi^2$ and the $\msbar$ renormalized top-quark mass,
$m_t$, are determined by~\cite{CEQR}
\be
\begin{array}{rl}
\label{runmasses}
m^2_H[\mu(t)]=& {\displaystyle  \frac{\xi^2[\mu(t^*)]}{\xi^2[\mu(t)]}
\left.\frac{\partial^2V_{\rm
eff}}{\partial\phi^2[\mu(t^*)]}\right|_{\phi[\mu(t^*)]=
\langle\phi[\mu(t^*)]\rangle}
},\vspace{0.3cm}\\
m_t[\mu(t)]= &{\displaystyle\frac{1}{\sqrt{2}}}h[\mu(t)]\xi[\mu(t)]v\ ,
\end{array}
\ee
and the pole masses $M_H$ and $M_t$ by~\cite{CEQ,Gray}
\be
\begin{array}{rl}
\label{polemasses}
M^2_H= & {\displaystyle m_H^2[\mu(t)]+{\rm
Re}\left[\Pi_{HH}(p^2=M_H^2)-\Pi_{HH}(p^2=0) \right] },\vspace{0.3cm}\\
M_t=& \left[1+{\displaystyle\frac{4}{3}\frac{\alpha_s(M_t)}{\pi}}\right] 
m_t[M_t] .\end{array}
\ee
It is clear that the effect of tadpole corrections is automatically 
included in both formulae. 

We can now comment on the errors associated with the estimates
of the pole masses (\ref{polemasses}). As for $M_t$, 
one-loop electroweak corrections (besides the tadpole ones) 
amount~\cite{HK} to $\sim$ 
+1\%, and the unconsidered two-loop QCD corrections~\cite{Gray} amount to
$\sim$ --1\%, so they almost cancel. In this way a conservative estimate 
of the total error
is $\sim$ 1\%, i.e. $\Delta M_t\simlt 2$ GeV. As for $M_H$, the
lack of scale invariance is a measure of the error associated
with the unconsidered corrections. This was studied in
Ref.~\cite{CEQ}, where an uncertainty of $\Delta M_H\simlt 2$
GeV was conservatively estimated.

It is clear from the above discussion that the claim in
Ref.~\cite{W2}, where it is argued that the discrepancy between AI and
CEQ results comes from the unconsidered large tadpole
corrections in Ref.~\cite{CEQ}, is incorrect. In fact, as we
have explained in this section, the large tadpole corrections
are absorbed in the one-loop VEV and should be nowhere in the
calculation relating the pole and $\msbar$ running masses,
either for the top-quark or the Higgs boson. 
The reason for the discrepancy should be traced back to the different
approximations used in the effective potential 
to compute the bounds in Refs.~\cite{AI} and \cite{CEQ}, and the
large uncertainty due to the lack of scale invariance
in the approximation of Ref.~\cite{AI}, 
as we will explain in the next section.

\section{The lower bound on $M_H$ and the origin of the discrepancy} 

It is well known that for certain values of the
top-quark and Higgs boson masses the effective potential
(\ref{potential}) develops an instability, i.e. the potential
becomes deeply negative, for large
values of the field~\cite{bounds}. Eventually, the potential
raises again yielding a (deep) non-standard minimum, although this
may happen for values of the field beyond the Planck scale.
To ensure that the electroweak minimum is the deepest one, 
the SM should be cutoff at a
value $\Lambda$ of the field such that the depth of the
potential equals the depth of the standard electroweak minimum.
For a fixed value of $\Lambda$ and $M_t$ this provides a lower
bound on $M_H$ such that the latter condition is barely
fulfilled. This procedure was recently refined in 
Refs.~\cite{AI,CEQ} where, as mentioned in the introduction, 
agreement was found, within the quoted
errors, for large values of $\Lambda$, while a large
discrepancy, impossible to reconcile within the quoted errors,
remains at low values of $\Lambda$. In this section we will
explain the origin of the discrepancy, reproduce the results of
Ref.~\cite{AI} and refine our previous analysis 
by controlling the scale invariance of the result.

The effective potential (\ref{potential}) improved by two-loop
renormalization group equations is highly scale independent
\cite{CEQR}. This allows fixing the renormalization scale as
$\mu(t)\sim \phi(t)$ in order to tame potentially dangerous 
logarithms at large values of the field \cite{effpot} (where the 
instability is expected to appear). In particular, fixing % 
\be
\label{fixscale}
\mu(t)=\alpha\phi(t)\; ,
\ee
allows to translate the scale-independence of the (whole) effective
potential into the $\alpha$ independence. Now, similarly
to the procedure 
followed in Ref.~\cite{CEQR} 
to determine $t^*$ for the standard
electroweak minimum, we can find out the optimum value of
$\alpha$ ($\alpha^*$ in what follows) to study the 
instability region
using the one-loop approximation to the potential [Eq.~(\ref{potential})]
and thus evaluating a more precise lower bound on $M_H$. The value of 
$\alpha^*$ will be 
that for which the results are more scale-invariant~\footnote{Again,
the precise value of $\alpha^*$ is
not very important due to 
the high degree of scale independence of the one-loop potential 
around $\alpha^*$, by the very definition of the latter.
In Ref.~\cite{CEQ} the reasonable choice $\alpha=1$ was made from
the very beginning.}.

Using (\ref{fixscale}) we can write the potential
(\ref{potential}) as
\be
\label{effpot}
V_{\rm eff}=-\frac{1}{2}m^2(t)\phi^2(t)+
\frac{1}{8}\lambda_{\rm eff}\phi^4(t) + \Omega(t)
\ee
where
\be
\label{efflambda}
\lambda_{\rm eff}(t)=\lambda(t)+
\sum_{i}\frac{n_i}{8\pi^2}\kappa_i^2\left[\log\frac{\kappa_i}{\alpha^2}
-C_i\right].
\ee
and $t=\log[\mu(t)/M_Z]$ is fixed by Eq.~(\ref{fixscale}).
The value of the scale $\Lambda$ where new physics has to stabilize the SM
potential is given by the value of the field $\phi$ where the
depth of the potential equals the depth of the potential at the
standard electroweak minimum. In practice, due to the steepness
of the potential around that point, we can identify
$\Lambda$ with the value of the field where the potential
vanishes, i.e.
\be
\label{condition}
\left.V_{\rm eff}(\phi)\right|_{\phi=\Lambda}= 0\;\;,
\ee
which can be written, using (\ref{effpot}) as
\be
\label{condlambda} 
\left.\left[\lambda_{\rm eff}-
4\frac{m^2}{\Lambda^2}
+8\frac{\Omega}{\Lambda^4}\right]\right|_{\mu=\alpha\Lambda}=0\;\;,
\ee
Since $\Omega$ obeys the one-loop RG equation \cite{effpot}
$8\pi^2 d \Omega/d t=m^4(t)$, it is clear that
for $\Lambda\simgt 1$ TeV its contribution to (\ref{condlambda})
is $\ll 1$~\footnote{The order of magnitude of $m(0)$ is
provided by the tree-level result $m^4\sim M_H^4/4$. Then, using
the boundary condition $\Omega(0)=0$ we obtain
$\Omega(\Lambda)\sim M_H^4/(32\pi^2)\log(\Lambda/M_Z)$.}, 
and thus can be compensated by a shift on $\lambda$ of the
same magnitude, with negligible consequences for the value of $M_H$.
Hence, the presence of $\Omega$ can be safely ignored. The 
$-4m^2/\Lambda^2$ term in (\ref{condlambda}) can have a small effect for 
low cut-offs being also negligible for larger ones. We include its 
effect in the numerical calculations although it is a very good 
approximation to write (\ref{condlambda}) as $\lambda_{\rm eff}\simeq 0$ for 
all values of the cut-off. In particular this will be assumed below 
when deriving some helpful analytical formulae.

In Fig.~1 we plot [thick solid line] the bound on $M_H$ as a function of
$\alpha$, for $\Lambda=1$ TeV,
$\alpha_s(M_Z)=0.118$ and $M_t=175$ GeV, from the condition
(\ref{condlambda}). Note that (\ref{condlambda}) depends on
$\alpha$ through Eq.~(\ref{fixscale}).
We see the mild dependence of $M_H$ on $\alpha$,
while the bound is flatter for the value $\alpha^*\sim 0.4$.
This behaviour was qualitatively expected from the analysis of
Ref.~\cite{CEQR} where the role of $\alpha^*$ here was
played there by the scale $t^*$ at the standard electroweak minimum.
In the same way that $t^*$ was an average of the masses appearing in the
one-loop corrections to the potential such that the effect of these 
corrections was cancelled, we have checked that $\alpha^*$ here can be 
estimated analytically with high precision as the value of $\alpha$ that 
cancels the one-loop corrections in Eq.~(\ref{efflambda}), which is
conceptually quite satisfactory.  
We recover, for $\alpha=1$, the bound $M_H\simgt 55$ GeV, in
agreement with our previous result in Ref.~\cite{CEQ}. 
Nevertheless, in the region where the bound is more scale independent
[i.e.  around $\alpha=\alpha^*$] we find $M_H\simgt 52$ GeV, which we
consider as the most reliable value. In particular, for
$\alpha^*/2\leq\alpha\leq 2\alpha^*$ we find $\Delta M_H\simlt 2$ GeV, 
which is a conservative estimate of the error~\footnote{Of 
course, going to too large or small values of $\alpha$ increases the size 
of logarithms in the perturbative expansion, and makes perturbation
theory unreliable. This phenomenon, expected from the point of
view of perturbation theory, was already observed with respect
to the $t^*$ dependence in Ref.~\cite{CEQR}.}.

We are now ready to discuss the origin of the discrepancy between the
AI and CEQ results. As we will see, the AI results are essentially
equivalent to a calculation based on 
the tree level potential improved by RG evolution at two-loops and thus 
resums all-loop leading and 
part of next-to-leading-logs. Our results are obtained
using the full one-loop potential with parameters running also at 
two-loops. This approximation for the potential 
resums all-loop leading and next-to-leading-logs and exhibits a high degree
of scale invariance, as was discussed in Ref.~\cite{CEQR}. In
fact, as shown there, the tree-level (leading-log) 
approximation exhibits a strong scale
dependence so that only provides a good approximation after a
judicious choice of the scale. Actually, we will explicitly show that
the results in Ref.~\cite{AI} are highly scale
dependent at low $\Lambda$ and can be made to 
coincide with the present results
for a particular value of the scale. Let us see this in somewhat more detail.

The scale $\Lambda$ is defined as the value of the $\phi$-field at
which the potential $V$ becomes negative. 
In principle, to study the value of $V$ at 
$\phi=\Lambda$ any value of the renormalization scale, $\mu$, could be used 
since the complete scalar potential is exactly scale invariant. 
But, when a perturbative approximation 
for the potential is used, exact scale invariance is lost and a convenient 
scale must be carefully chosen.
Although the authors of Ref.~\cite{AI} write in principle
the complete one-loop potential, their criterion, in practice, 
to identify $\Lambda$ as the value of the scale at which
$\lambda$ crosses zero, namely $\lambda(\Lambda)=0$, is essentially 
equivalent to use $\mu=\Lambda$ [i.e. $\alpha=1$ in 
Eq.~(\ref{fixscale})] and require $V_0(\phi=\Lambda)=0$;
or, in other words, to ignore~\footnote{
As we will see, this is in fact a good approximation for large values of
$\Lambda$, but not for the low $\Lambda$ regime, in which we are concerned
here.} the one-loop corrections to 
${\lambda}_{\rm eff}$ given by (\ref{efflambda}).
However, although it is clear that an appropriate
scale to evaluate the potential should be of the same order
as the value of the field in which we are interested, 
there is no reason to expect them to be 
exactly equal. This consideration is particularly relevant for the AI
results since the (RG improved) potential $V_0$ is strongly dependent
on the renormalization scale, and so are the corresponding results on 
$M_H$. This can be seen from Fig.~1 where we have plotted the 
corresponding bound on $M_H$ as a function of $\alpha$, i.e. the
renormalization scale [thin solid line].
As it is clear from the figure, in this approximation
the bound is, as expected, strongly dependent on $\alpha$. 
For $\alpha=1$ we obtain the bound $M_H\simgt 72$ GeV, which is
the AI bound quoted in Ref.~\cite{AI}. For $\alpha\sim\alpha^*$, the
two approximations coincide since, as mentioned above, for that scale 
the radiative corrections to $V_0$ are essentially 
cancelled~\footnote{A similar {\it coincidence} was also
observed in Ref.~\cite{CEQR} concerning the scale $t^*$.}.
Therefore, for a given value of $M_H$ the value for $\Lambda$
obtained  by CEQ and the one obtained by AI (say $\Lambda_0$) are
related by 
\be
\label{relescalas}
\Lambda_0\equiv\alpha^*\Lambda=
\exp\left\{\Delta\lambda(\Lambda_0)/\beta_{\lambda}(\Lambda_0)
\right\}\Lambda\; ,
\ee
where we have used the scale invariance of the effective potential to
obtain the second equality, and~\footnote{Notice that 
$\Delta\lambda(\Lambda_0)$ has a clear  
interpretation. If we fix the stability scale $\Lambda$ and define
two different boundary conditions (Higgs masses) for $\lambda$: the 
AI boundary condition [i.e. $\lambda_{\rm AI}(\Lambda)=0$] and the
CEQ boundary condition [i.e. eq.(\ref{condlambda}), or equivalently
$\lambda_{\rm CEQ}(\Lambda_0)\simeq 0$];
then, $\Delta\lambda(\Lambda_0)\simeq\lambda_{\rm AI}(\Lambda_0)
-\lambda_{\rm CEQ}(\Lambda_0)$.}

\be
\begin{array}{rl}
\Delta\lambda(\Lambda_0)= & {\displaystyle
\sum_{i}\frac{n_i}{8\pi^2}\kappa_i^2\left[\log\kappa_i-C_i\right] }\\
\beta_{\lambda}(\Lambda_0)= & {\displaystyle
\frac{1}{16\pi^2}\left[-12h_t^4+\frac{9}{4}
\left(g^4+\frac{2}{3}g^2g'^2+\frac{1}{3}g'^4\right)\right] }\; .
\end{array}
\label{deltas}
\ee

For illustrative purposes we have plotted in Fig.~2 the
effective potential (\ref{effpot}) [thick solid line],
for $\alpha=\alpha^*$, $\Lambda=1$ TeV, 
$\alpha_s(M_Z)=0.118$, $M_t=175$ GeV and
$M_H=52$ GeV. We see that the value of the field $\phi$ where
the potential satisfies condition (\ref{condition}) and where
the value of the potential equals the depth of the standard
electroweak minimum are almost indistinguishable, as anticipated. We
also plot $\lambda_{\rm eff}(\alpha^*\phi)$ 
[thick dashed line] as a function of the
field and subject to the boundary condition 
$\lambda_{\rm eff}(\alpha^*\Lambda)=0$. As 
expected, $\lambda_{\rm eff}$ crosses zero exactly 
at the value of the field $\phi$ where the potential
itself crosses zero so that it keeps track perfectly of the potential 
destabilization. The plot of $\lambda(\phi)$ 
[thin dashed line] crosses zero
at the scale $\Lambda_0\sim 370$ GeV, in good agreement with
Eq.~(\ref{relescalas}). In other words, had we used the tree-level condition
$\lambda(\Lambda)=0$, we would have obtained the same value for the
Higgs mass lower bound, i.e. 52 GeV, but for the scale $\sim$ 370 GeV.

Let us now discuss why the AI results are in good agreement
with the CEQ ones for large values of $\Lambda$. This can be intuitively
understood from the very small dependence of the $M_H$ bound on $\Lambda$
for large $\Lambda$ (see either Refs. \cite{AI} or \cite{CEQ}). Then, it
is clear that the uncertainty derived from the choice of the scale in the 
AI approximation  becomes very small and hence
AI and CEQ results get agreement. A more precise way to understand the
agreement is to note that $\Delta\lambda(\Lambda_0)$ 
gets reduced for large scales because the top Yukawa 
coupling runs to smaller values at high scales and, besides, there is
a cancellation between the top Yukawa and gauge effects~\footnote{
The value of $\beta_\lambda$ is also much smaller for the same reasons,
producing the above mentioned insensitivity of $M_H$ to the value
of $\Lambda$ for high $\Lambda$.} in $\Delta\lambda(\Lambda_0)$.
On top of that, the fact that $M_H$ is larger for higher scales contributes
to the coincidence between AI and CEQ results. Therefore, for a 
given discrepancy $\delta M_H^2$, the effect in 
$\delta M_H\sim\delta M_H^2/2 M_H$ is suppressed as $M_H$ grows.

To illustrate these effects we present in Table 1 the values of $\alpha^*$ 
[i.e. $\Lambda_0$], the bounds on $M_H$, and the couplings $g$, $g'$ and $h_t$ 
evaluated at the scale $\Lambda_0$ for two different sets of parameters: the 
case discussed in Fig.~2, $\Lambda=1$ 
TeV, and the case of a high value of $\Lambda$, in particular
$\Lambda=10^{19}$ GeV. 
\begin{center}
\begin{tabular}{||c|c|c|c|c|c||}\hline
$\Lambda$[GeV] & $\Lambda_0$[GeV]&$M_H$[GeV]& $g'$ & $g$ & $h_t$ \\ \hline
% & & & & & \\
$10^3$ & 370 & 52 & 0.358 & 0.643 & 0.912 \\ \hline
% & & & & & \\
$10^{19}$& $3.6\times 10^{17}$ & 134& 0.457 & 0.514& 0.414 \\ \hline
\end{tabular}

\vspace{0.5cm}
Table 1
\end{center}
The squared mass difference between 
AI and CEQ calculations can be approximated by (see footnote 7)
\be
\label{massdiff}
\delta M_H^2=\Delta\lambda(M_t)v^2=\Delta\lambda(\Lambda_0)v^2
+\cdots
\ee
where $\Delta\lambda(\Lambda_0)$ is defined in Eq.~(\ref{deltas}) and
the ellipsis stands for the renormalization of $\Delta\lambda$ from
$\Lambda_0$ to $M_t$, which is a small effect in all cases. Then,
using Table 1 and Eq.~(\ref{deltas}) one can obtain,
disregarding radiative corrections as in (\ref{massdiff}),
for $\Lambda=1$ TeV, $\delta M_H^2\sim (60\ {\rm GeV})^2$, and
for $\Lambda=10^{19}$ GeV, $\delta M_H^2\sim (10\ {\rm GeV})^2$, which
explains qualitatively the agreement (disagreement) between AI and CEQ
for large (small) values of $\Lambda$.

To close the discussion on
the bound for large cut-offs we would also like to comment on the results
of Ref.~\cite{W1}. Although it is clear that the recipe given in that paper to 
compute the bound is exactly the same used by AI, somewhat larger bounds
are found for the case of $\Lambda=M_P$. Part of this effect can be 
explained by the fact that \cite{W1} integrates one-loop RGEs while AI
is using two-loop equations instead. As is well known \cite{AI}, this will 
result in an overestimated bound, the effect being more important the longer
the running is.

\section{Detailed results and estimate of errors}

In Fig.~3 we have plotted the lower bounds on $M_H$ based on
condition (\ref{condlambda}) [solid lines], for $\Lambda=1$ TeV
and $\alpha_s(M_Z)=0.118\pm 0.006$,
as functions of $M_t$. A very accurate fit (with an error below
1~GeV), is given by
\be
\label{fit}
M_H[GeV]>52+0.64\ (M_t[GeV]-175)-0.50\ \frac{\alpha_s(M_Z)-0.118}{0.006}.
\ee
For the sake of comparison we also plot [dashed line] the
corresponding bound based on the condition $\lambda(\Lambda)=0$,
for $\alpha_s(M_Z)=0.118$. We can
see that its prediction agrees well with AI bounds in
Ref.~\cite{AI}. In Fig.~4 we plot the lower bounds based on our
condition (\ref{condlambda}) as functions of $\Lambda$ for
values $M_t=$ 150, 175 and 200 GeV,
and $\alpha_s(M_Z)=0.118$. For a fixed value of $M_t$
the corresponding curve provides an upper bound on the scale of
new physics necessary to stabilize the SM potential. This bound
can be read from Fig.~4 as a function of $M_H$. 
A very good fit is provided by
\be
\label{fitlambda}
\log\Lambda[TeV]<a_0+a_1 x+a_2 x^2,
\ee
where
\be
\label{exis}
x=\frac{M_H[GeV]-75}{10},
\ee
and $a_i$ (i=0,1,2) are given in Table 2.

\begin{center}
\begin{tabular}{||c|c|c|c||}\hline
$M_t$[GeV] & $a_0$ & $a_1$ & $a_2$ \\ \hline
150 & 4.62 & 1.84 & 0.17 \\ \hline
175 & 1.39 & 0.76 & 0.08 \\ \hline
200 & 0.24 & 0.36 & 0.04 \\ \hline
\end{tabular}

\vspace{0.5cm}
Table 2
\end{center}

We will conclude by making an evaluation of the errors affecting
our analysis. First of all, those from the determination of the pole
masses for the top-quark and Higgs boson (\ref{polemasses}) have
already been mentioned in section 2. We
have seen that unconsidered two-loop QCD and one-loop
electroweak corrections to $M_t$ lead to an uncertainty in $M_t$
of $\sim$ 1\%, i.e. to $\Delta M_t\simlt 2$ GeV, which
translates from (\ref{fit}) to $\Delta M_H\sim$ 1 GeV. On the
other hand, the scale dependence of $M_H$, which measures the
unconsidered higher-order corrections, was evaluated in
Ref.~\cite{CEQR} to $\Delta M_H\sim$ 2 GeV. The other source of
theoretical uncertainty comes from the lack of exact
renormalization-scale invariance of the one-loop effective
potential at the instability region (which has been encoded into the 
parameter $\alpha$). This (mild) scale dependence reflects all the 
unconsidered higher-order effects, in particular higher-loop corrections,
thus being a good measure of the theoretical uncertainty of the calculation.
As it was stated in the previous section, a variation of $\alpha$
in the range $[\alpha^*/2, 2\alpha^*]$ yields
$\Delta M_H\simlt 2$ GeV, which we
consider to be a conservative estimate of the theoretical
uncertainty [values of $\alpha$ far from the `optimum'
value, $\alpha^*$, lead to large
logarithms in the perturbative expansion, thus becoming unreliable,
see footnote 4]. Note that the value of $\alpha^*$ we are using
corresponds to the choice that cancels one-loop corrections
(including finite contributions). This way of fixing $\alpha^*$
has sometimes the drawback that higher order logarithms
$[\log(\kappa_i/\alpha^2)]^n$ are not automatically zero
and can be enventually important.
It would be nice to confirm this by the explicit 
computation of the next higher order corrections
to our calculation. This in fact can be done by adding the 
leading ${\cal O}(g_s^2 h_t^4)$ and
${\cal O}(h_t^6)$ two-loop effective potential
corrections (the full two-loop potential can be found in Ref.~\cite{FJJ}), 
which provide a two-loop 
correction to the effective coupling $\lambda_{\rm eff}$ given by
\bea
\label{twocorr}
\Delta_{\rm 2-loop}\lambda_{\rm eff}
&= 
{\displaystyle\frac{2 h_t^4}{(16\pi^2)^2}}
\left\{g_s^2\left[24\left(\log{\displaystyle\frac{h_t^2}{2\alpha^2}}
\right)^2
-64\log{\displaystyle\frac{h_t^2}{2\alpha^2}}+72\right]
 \right. \nonumber \vspace{0.3cm}\\
& 
\left.-{\displaystyle\frac{3}{2} }h_t^2
\left[3\left(\log{\displaystyle\frac{h_t^2}{2\alpha^2}}
\right)^2
-16\log{\displaystyle\frac{h_t^2}{2\alpha^2}}
+23+{\displaystyle\frac{\pi^2}{3} }\right]\right\}.
\eea
This correction (which is positive for any value of $\alpha$) 
raises the value of ${\lambda}_{\rm eff}$ if the boundary condition
of $\lambda$ is maintained. Therefore, condition (\ref{condlambda})
indicates that for a given $\Lambda$, the boundary condition of $\lambda$
must actually be slightly lowered, and thus the $M_H$ bound.
For the typical case $\Lambda=1$ TeV, $M_t=175$ GeV, we find
$\Delta M_H\sim -1$ GeV,
%
%~\footnote{More precisely, defining $\alpha^*$
%as the value of $\alpha$ where the one-loop radiative corrections in
%Eq.~(\ref{efflambda}) vanish, and by $\alpha^*+\Delta\alpha^*$ the value of
%$\alpha$ where the one- plus two-loop radiative corrections, after the
%inclusion of Eq.~(\ref{twocorr}) in (\ref{efflambda}) vanish, we obtain
%$\Delta\alpha^*\sim -0.01$, which would correspond to the same shift 
%in the crossing point in Fig.~1 between the thick and thin solid lines.
%This shift amounts, as can be read off from Fig.~1, to 
%$\Delta M_H\sim -1$ GeV.}, 
%
well within the previous conservative
estimate $\Delta M_H\simlt 2$ GeV.

Let us now discuss the dependence of our bounds on the renormalization 
scheme. In principle one could expect small changes associated
with the choice of the scheme. However, since an exact calculation 
would remove all the ambiguities as well as the scale dependence for 
any physical quantity, assigning an additional error to this effect 
would be redundant. It is illustrative to comment how this works
when one compares the results obtained in the MS and $\msbar$ 
schemes. For them, the change in the 
definition of $\lambda_{\rm eff}$ in Eq.~(\ref{efflambda}), and the
subsequent modification in condition (\ref{condlambda}), is exactly
compensated at one loop by the modification in the relation to extract
the value of $M_H$, Eqs.~(\ref{runmasses}), (\ref{polemasses}).
This is not surprising since the relation between MS and $\msbar$ 
schemes can be viewed as a redefinition of the renormalization scale,
and our calculation is scale invariant up to higher order corrections.
However, the use of the tree-level condition $\lambda(\Lambda)=0$
does produce different results in the MS and $\msbar$ schemes due to
the lack of scale invariance of the approach.

Concerning the gauge dependence of the bounds, it is expected to be well
below the previously estimated errors. Note that the main effect of loop
corrections in the effective potential
can be assigned to the Yukawa sector where no gauge dependence
arises. In fact, including the one-loop gauge corrections 
(as we have done numerically) amounts to shifting the bound by
$\Delta M_H\sim 0.5$ GeV, a negligible quantity as compared to our
previously estimated errors

Altogether  we can
conclude that a realistic evaluation on the error associated
with our calculational method yields an uncertainty of
$$\Delta_{\rm tot} M_H\simlt 5\ GeV,$$ 
which can be taken as a conservative estimate. 
This error does not take
into account those coming from the measured values of $M_t$ and
$\alpha_s(M_Z)$. For $\Lambda=1$ TeV, the former can be readily 
read off from Eq.~(\ref{fit}) to be given by 
$$\Delta M_H=0.64\ \Delta M_t$$
while the latter (much smaller) is estimated to be
$$\Delta M_H\sim -0.50\ \frac{\Delta\alpha_s(M_Z)}{0.006}.$$

\section{Conclusions}

In this letter we have clarified some
recent controversy concerning stability bounds on the SM Higgs
mass for low values of the SM cutoff, $\Lambda$. 
We have reanalyzed our previous results by making a refined
analysis of the scale dependence of the effective potential,
which these bounds depend upon, reobtained other results
existing in the literature and shown that the latter can be
explained as a consequence of the corresponding approximations 
made in evaluating the effective potential. We have restricted
ourselves to scales ($\Lambda\simlt 10$ TeV) which are within
the reach of LHC. For that reason our detailed
analysis is relevant for the phenomenology of the planned
colliders. In particular, from Fig.~4 we see that for $M_t\simgt
175$ GeV, if the Higgs is found at LEP2 
with a mass  $M_H\simlt 85$ GeV, 
then $\Lambda\simlt 10$ TeV, which implies that new physics
(if weakly coupled),
necessary to stabilize the SM effective potential, should be
produced at LHC. Even if we admit living in a metastable
minimum, with a lifetime to the non-standard minimum longer
than the present age of the universe, the detailed bounds,
obtained from the calculation of the decay to the non-standard
minimum~\cite{EQ}, as a function of $\Lambda$ do depend on the
very existence and location of the latter, and thus a detailed 
knowledge of it is still relevant.

\section*{Acknowledgements}

We thank G.~Altarelli, M.~Carena, H.~Haber, C.~Wagner and 
F.~Zwirner for discussions and useful comments.

\newpage
%%%%%%%%%%%%%%%%%%%%%%%%figure%%%%%%%%%%%%%%%%%%%%%%%%
\begin{figure}
%\psdraft
\centerline{
\psfig{figure=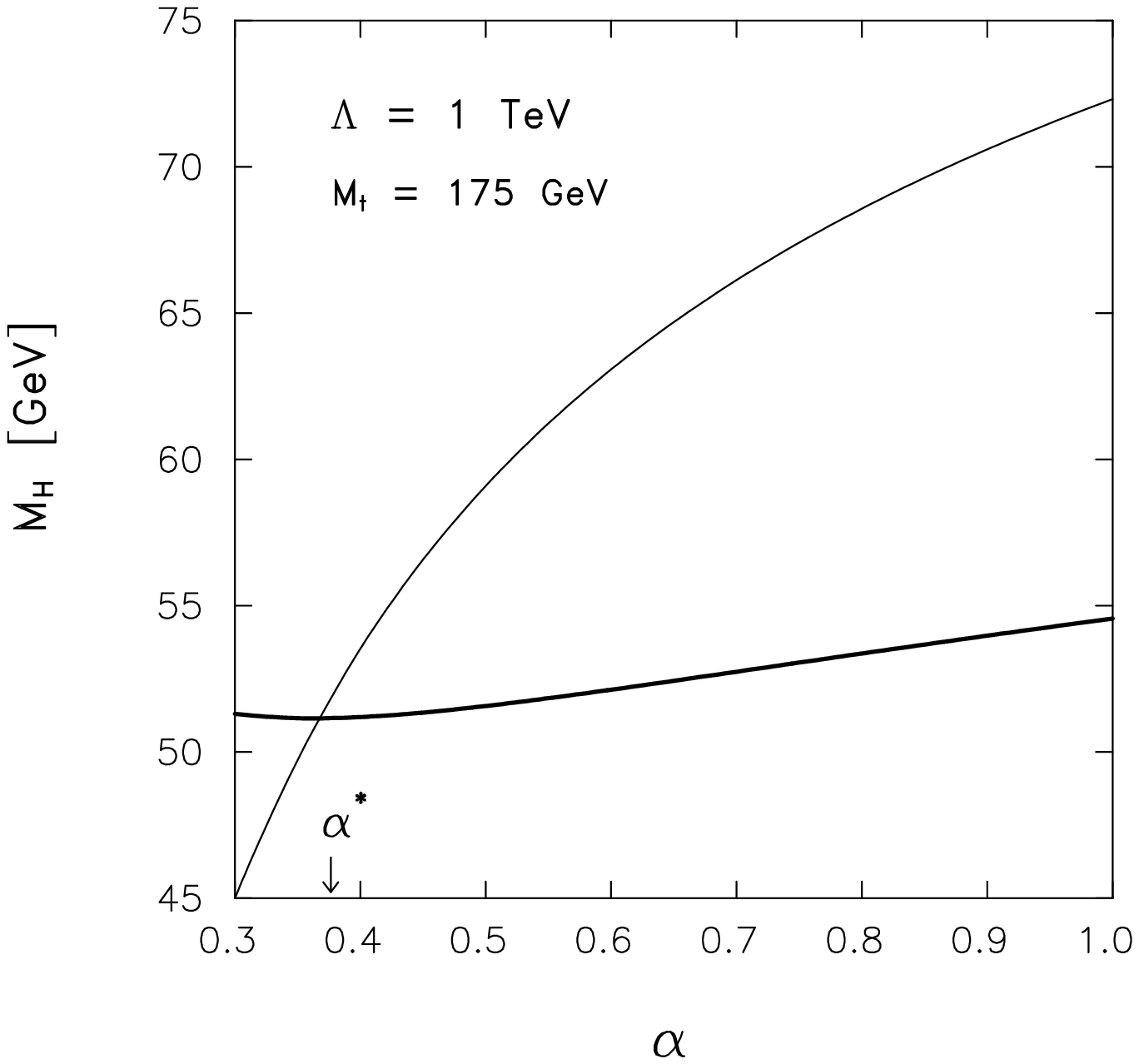,height=13cm,bbllx=4.5cm,bblly=.cm,bburx=14.cm,bbury=13cm}}
\caption{Plots of the lower bound on the Higgs mass based on 
condition (15) [thick solid line], and on the
condition $\lambda(\alpha\Lambda)=0$ [thin solid line], 
as functions of the parameter $\alpha$, defined by $\mu(t)=\alpha\phi(t)$, 
for $\Lambda=1$ TeV, $\alpha_s(M_Z)=0.118$ and $M_t=175$ GeV.}

\end{figure}
%%%%%%%%%%%%%%%%%%%%%%%%figure%%%%%%%%%%%%%%%%%%%%%%%%
\begin{figure}
%\psdraft
\centerline{
\psfig{figure=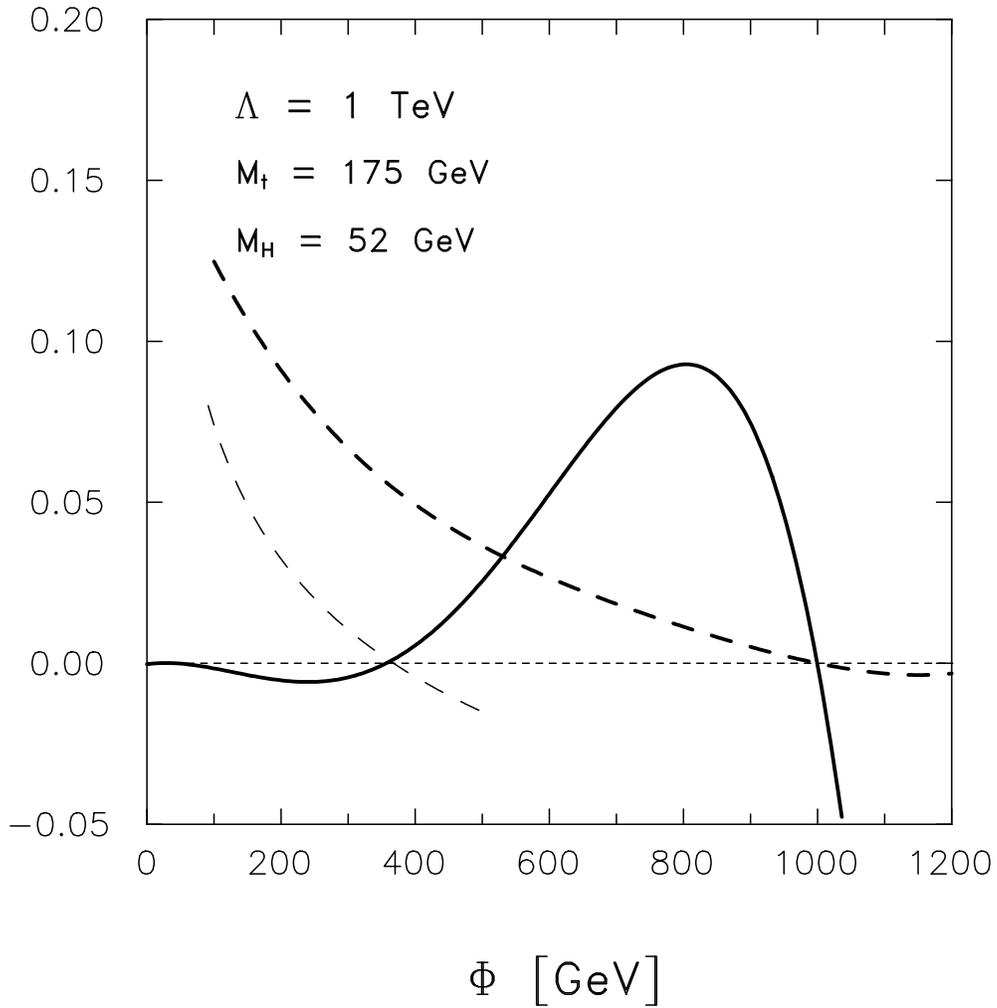,height=13cm,bbllx=4.5cm,bblly=.cm,bburx=14.cm,bbury=13cm}}
\caption{Plot of the effective potential [thick
solid line] for $\Lambda$, $\alpha_s(M_Z)$ and $M_t$ as in Fig.~1, 
and $M_H=52$ GeV.
Dashed lines are plots of $\lambda_{\rm eff}(\mu=\alpha^*\phi)$ [thick one]
and $\lambda(\mu=\phi)$ [thin one].}
\end{figure}
%%%%%%%%%%%%%%%%%%%%%%%%figure%%%%%%%%%%%%%%%%%%%%%%%%
\begin{figure}
%\psdraft
\centerline{
\psfig{figure=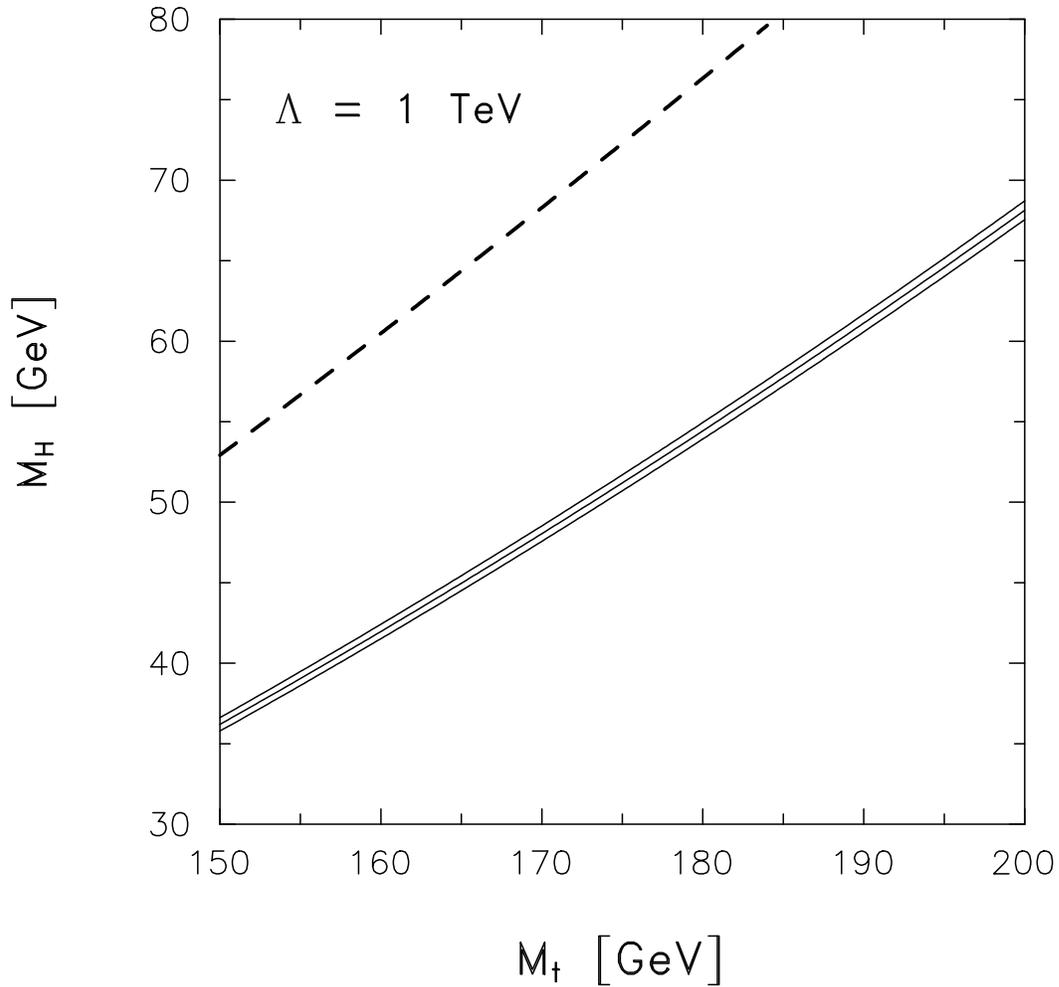,height=13cm,bbllx=4.5cm,bblly=.cm,bburx=14.cm,bbury=13cm}}
\caption{Plots of the lower bound on $M_H$  as a
function of $M_t$ from condition
$\lambda_{\rm eff}=0$
[solid lines] as in Fig.~1, for $\alpha_s(M_Z)=0.118$ [central line],
$\alpha_s(M_Z)=0.124$ [lower line] and $\alpha_s(M_Z)=0.112$ [upper line].
The bound based on the
condition $\lambda(\Lambda)=0$ and $\alpha_s(M_Z)=0.118$
is also plotted for the sake of
comparison [dashed line].}
\end{figure}
%%%%%%%%%%%%%%%%%%%%%%%%%%figure%%%%%%%%%%%%%%%%%%%%%%%%%%%%
\begin{figure}
%\psdraft
\centerline{
\psfig{figure=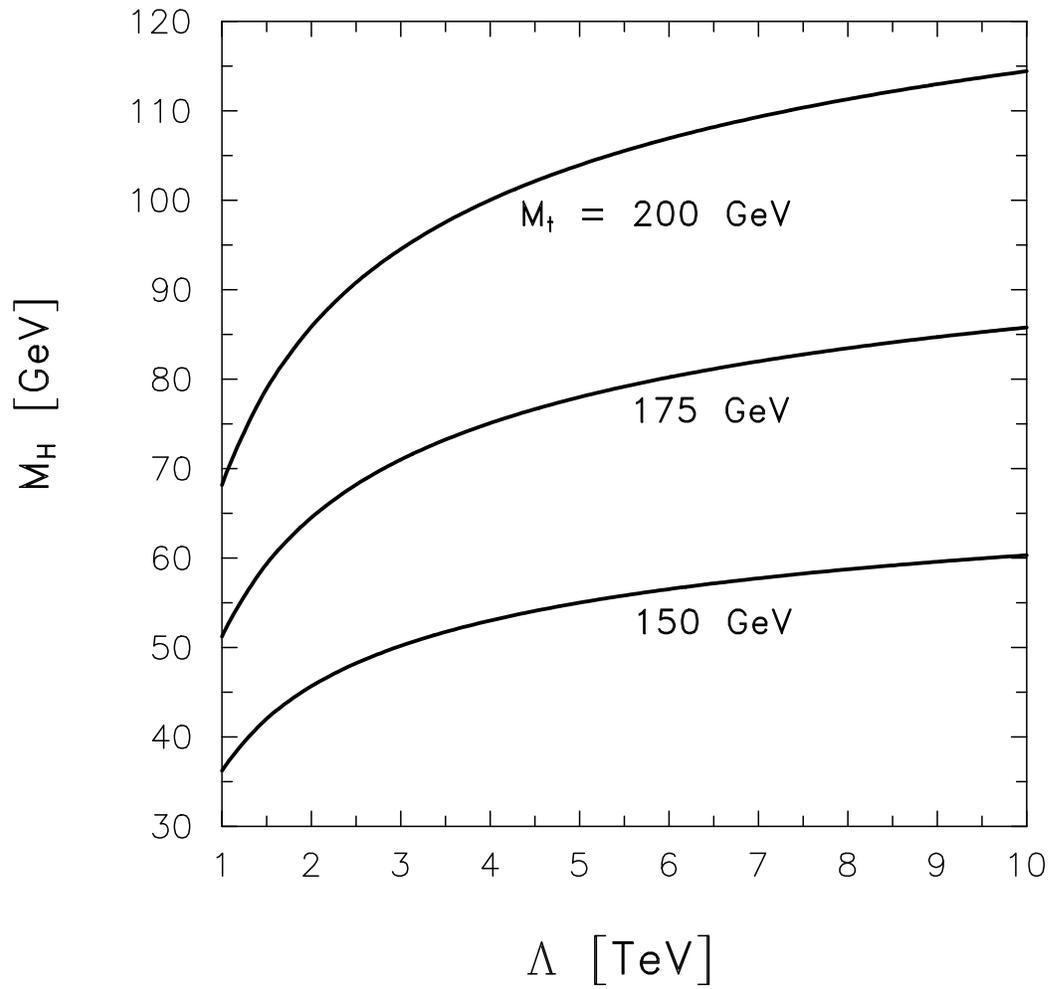,height=13cm,bbllx=4.5cm,bblly=.cm,bburx=14.cm,bbury=13cm}}
\caption{Plots of the lower bound on $M_H$ as a function of
$\Lambda$ for different values of $M_t$ and $\alpha_s(M_Z)=0.118$.}
\end{figure}
%%%%%%%%%%%%%%%%%%%%%%%%figure%%%%%%%%%%%%%%%%%%%%%%%%


\newpage
\begin{thebibliography}{99}
%
\bibitem{bounds}
N. Cabibbo, L. Maiani, G. Parisi and R. Petronzio, \NPB{158}{79}{295};
M.J. Duncan, R. Phillipe and M. Sher,
\PLB{153}{85}{165}; M.~Lindner, \ZPC{31}{86}{295};
M.~Sher, \PR{179}{89}{273};
M.~Lindner, M.~Sher and H.W.~Zaglauer, \PLB{228}{89}{139};
M.~Sher, \PLB{317}{93}{159}; Addendum:  \PLB{331}{94}{448} 
%
\bibitem{HS} P.Q.~Hung and M.~Sher, preprint WM-95-111
[hep-ph/9512313] 
%
\bibitem{AI} G.~Altarelli and I.~Isidori, \PLB{337}{94}{141}
%
\bibitem{CEQ} J.A.~Casas, J.R.~Espinosa and M.~Quir\'os, 
\PLB{342}{95}{171}
%
\bibitem{W1} R.S.~Willey, Pittsburg preprint [hep-ph/9512226]
%
\bibitem{W2} R.S.~Willey, Pittsburg preprint [hep-ph/9512286]
%
\bibitem{effpot} B.~Kastening, \PLB{283}{92}{287};
C.~Ford, D.R.T.~Jones, P.W.~Stephenson and 
M.B.~Einhorn, \NPB{395}{93}{17};
M.~Bando, T.~Kugo, N.~Maekawa and H.~Nakano, \PLB{301}{93}{83};
\PTP{90}{93}{405}
%
\bibitem{CEQR} J.A.~Casas, J.R.~Espinosa, M.~Quir\'os and A.~Riotto,
\NPB{436}{95}{3}
%
\bibitem{HK} R.~Hempfling and B.A.~Kniehl, \PRD{51}{95}{1386}
%
\bibitem{BW} A.I.~Bochkarev and R.S.~Willey, \PRD{51}{95}{R2049}
%
\bibitem{Gray} N.~Gray, D.J.~Broadhurst, W.~Grafe and K.~Schilcher,
\ZPC{48}{90}{673}
%
\bibitem{FJJ} C.~Ford, I.~Jack and D.R.T.~Jones, \NPB{387}{92}{373}
%
\bibitem{EQ} J.R.~Espinosa and M.~Quir\'os, \PLB{353}{95}{257}
%
\end{thebibliography}
\end{document}